\documentclass[aps,pra,reprint,floatfix,nobibnotes,longbibliography]{revtex4-1}
\pdfoutput=1
\usepackage{amsmath}
\usepackage[version=4]{mhchem}
\usepackage{xcolor}
\usepackage{units}
\usepackage{graphicx}

\newcommand{\voc}{V_\text{oc}}
\newcommand{\vocmax}{V_\text{oc,max}}
\newcommand{\vocone}{V_{\text{oc},1}}
\newcommand{\voctwo}{V_{\text{oc},2}}

\newcommand{\vbi}{V_\text{bi}}

\newcommand{\js}{j_S}
\newcommand{\nmin}{n_\text{min}}
\newcommand{\mun}{\mu_n}
\newcommand{\mup}{\mu_p}
\newcommand{\mueff}{\mu_\text{eff}}
\newcommand{\sigmap}{\sigma_p}
\newcommand{\sigman}{\sigma_n}
\newcommand{\eg}{E_g}
\newcommand{\efn}{E_{F,n}}
\newcommand{\efp}{E_{F,p}}
\newcommand{\phian}{\varphi_\text{an}}
\newcommand{\phicat}{\varphi_\text{cat}}
\newcommand{\moox}{MoO$_x$}
\newcommand{\pedot}{PEDOT:PSS}

\begin{document}

\title{Effect of Imbalanced Charge Transport on the Interplay of Surface and Bulk Recombination in Organic Solar Cells}

\author{Dorothea Scheunemann}
\email[]{dorothea.scheunemann@uol.de}
\author{Sebastian Wilken}
\affiliation{Physics, Faculty of Science and Engineering and Center for Functional Materials, \AA{}bo Akademi University, Porthansgatan 3, 20500 Turku, Finland}
\affiliation{Institute of Physics, Energy and Semiconductor Research Laboratory, Carl von Ossietzky University of Oldenburg, 26111 Oldenburg, Germany}

\author{Oskar J.~Sandberg}
\affiliation{Department of Physics, Swansea University, Singleton Park, Swansea, SA2 8PP, Wales, United Kingdom}

\author{Ronald \"{O}sterbacka}
\affiliation{Physics, Faculty of Science and Engineering and Center for Functional Materials, \AA{}bo Akademi University, Porthansgatan 3, 20500 Turku, Finland}

\author{Manuela Schiek}
\affiliation{Institute of Physics, Energy and Semiconductor Research Laboratory, Carl von Ossietzky University of Oldenburg, 26111 Oldenburg, Germany}

\begin{abstract}
Surface recombination has a major impact on the open-circuit voltage~($\voc$) of organic photovoltaics. Here, we study how this loss mechanism is influenced by imbalanced charge transport in the photoactive layer. As a model system, we use organic solar cells with a two orders of magnitude higher electron than hole mobility. We find that small variations in the work function of the anode have a strong effect on the light intensity dependence of~$\voc$. Transient measurements and drift-diffusion simulations reveal that this is due to a change in the surface recombination rather than the bulk recombination. We use our numerical model to generalize these findings and determine under which circumstances the effect of contacts is stronger or weaker compared to the idealized case of balanced charge transport. Finally, we derive analytical expressions for~$\voc$ in the case that a pile-up of space charge is present due to highly imbalanced mobilities.
\end{abstract}

\maketitle

\section{Introduction}
\label{sec:intro}
Organic solar cells typically consist of a blend of an electron donor and an electron acceptor sandwiched between two electrical contacts~\cite{Deibel2010,Mishra2012,Hou2018}. Ideally, the contacts act as semipermeable membranes for electrons~(cathode) and holes~(anode). In this case, the open-circuit voltage $\voc$ is solely determined by the splitting of the quasi-Fermi levels $\efn - \efp$ in the blend~\cite{Wurfel2009,Vandewal2009}. If only bimolecular recombination is present, 
\begin{equation}
q \voc = \eg - kT \ln\left(\frac{\beta N^2}{G}\right)
\label{eq:koster}
\end{equation}
has been suggested, where $q$ is the elementary charge, $\eg$ the band gap, $k$ the Boltzmann constant, $T$ the temperature, $\beta$ the recombination coefficient, $N$ the density of states and $G$ the generation rate~\cite{Koster2005,Tress2012}. The latter part of Eq.~(\ref{eq:koster}) predicts a slope of $kT/q$ when plotting $\voc$ versus the logarithm of the light intensity.

However, many contacts like metals or doped polymers are non-selective contacts. This means they have the ability to exchange both minority and majority charge carriers with the photoactive layer. The extraction of minority carriers (electrons at the anode, holes at the cathode) leads to a reduction of $\voc$~\cite{Reinhardt2014}. Here, we call this loss mechanism surface recombination, with a corresponding recombination current
\begin{equation}
\js = q S \nmin,
\label{eq:js}
\end{equation}
where $S$ denotes the surface recombination velocity and $\nmin$ the concentration of minority carriers close to the contact under consideration~\cite{Wagenpfahl2010,Sandberg2014}.

In the case of Ohmic contacts, $\js$ is strongly reduced by charge carrier injection. Because of the high concentration of majority carriers at the interface, minority carriers are much more likely to recombine in the bulk, rather than leaving the device via the ``wrong'' electrode. As a result, $\voc$ is still determined solely by properties of the bulk and follows Eq.~(\ref{eq:koster}).

The situation changes when one of the contacts is non-Ohmic. For instance, if an injection barrier $\phian$ is present at the anode, less holes are injected into the blend, so that $\nmin$ and $\js$ are effectively increased. Solak~et~al.~\cite{Solak2016} showed that the open-circuit voltage at high light intensities may then be described by
\begin{equation}
q \voc = \eg - \phian - \frac{kT}{2} \ln\left(\frac{\beta N^2}{G}\right).
\label{eq:solak}
\end{equation}
Compared to Eq.~(\ref{eq:koster}), there are two differences: First, the constant energetic part is reduced by the barrier height. Second, because of the factor~$1/2$ in front of the logarithm, the intensity dependence of $\voc$ is now given by a slope of $kT/2q$ (instead of $kT/q$). Such a reduction of the slope has been demonstrated both in experiment and simulation~\cite{Sandberg2016,Wheeler2015,Solak2016}. The transition between Eq.~(\ref{eq:koster}) and Eq.~(\ref{eq:solak}) has been assumed to take place when $\voc$ equals the built-in voltage $\vbi$~\cite{Solak2016}. 

For large $S$, surface recombination is not limited by the interface kinetics, but the transport of carriers towards the contact~\cite{Sandberg2016}. Hence, the question arises, how the open-circuit voltage depends on the charge carrier mobility~$\mu$. Numerical studies have indicated a decrease of $\voc(\mu)$ with increasing mobility if the contacts are non-selective~\cite{Tress2012,Kirchartz2009,Wagenpfahl2010b}. The result is a finite optimum value of $\mu$ in terms of the overall device efficiency, independent of the recombination mechanism in the bulk.

In the above considerations, the mobilities of electrons~($\mun$) and holes~($\mup$) are considered balanced. However, this condition is often not fulfilled in practice. Many polymer-fullerene solar cells, for instance, exhibit a higher electron than hole mobility~\cite{Bartelt2015,Stolterfoht2016}, while it is the other way round for devices based on recent non-fullerene acceptors~\cite{Holliday2016,Yan2018,Hou2018}. It is well known that the imbalanced mobilities lead to a pile-up of space charge close to one contact, which may reduce both the fill factor and short-circuit current~\cite{Mihailetchi2005,Stolterfoht2015}. In contrast, little attention has been paid on how this affects the open-circuit voltage. Recently, Spies~et~al.~\cite{Spies2017} suggested that the additional charge will further reduce the built-in potential and, thus, severely affect the magnitude of $\voc$.

In this work, we use an experimental system with a strong mobility mismatch of $\mun/\mup = 100$ and a well calibrated numerical model to discuss the effect of imbalanced transport on the open-circuit voltage in more detail. We show that the ratio between electron and hole mobility critically determines whether~$\voc$ is dominated by surface recombination or bimolecular recombination in the bulk. With the help of the numerical simulations we expand the analytical framework given by Eq.~(\ref{eq:solak}) to the case of imbalanced mobilities.

\section{Experimental and Numerical Framework}
\label{sec:exp}

\subsection{Experiment}
We fabricated solar cells based on a bulk heterojunction of the small molecule donor 2,4-bis[4-(\textit{N,N}-di\-iso\-butyl\-amino)-2,6-di\-hydroxy\-phenyl] squaraine~(SQIB) and the fullerene acceptor [6,6]phenyl-C$_{61}$-butyric acid methyl ester~(PCBM). This blend is known for a strong contrast between the mobility of electrons~($\unit[10^{-4}]{cm^2/Vs}$) and holes~($\unit[10^{-6}]{cm^2/Vs}$)~\cite{Scheunemann2017}. Our devices had the structure indium tin oxide/HTL/SQIB:PCBM/LiF/Al, where HTL denotes the hole transport layer. Details regarding the used materials and the device preparation can be found elsewhere~\cite{Bruck2014,Abdullaeva2016,Scheunemann2017}.

To realize both devices with an Ohmic and a non-Ohmic contact, we changed the HTL from molybdenum suboxide~(\moox) to the doped polymer~\pedot. The energy level of~\pedot\ lies within the band gap of the photoactive blend, effectively reducing $\vbi$ by \unit[280]{mV} (see Supplemental Material~\cite{SM}). As shown in Fig.~\ref{fig:figure01}, this reduction results primarily in a significant drop of $\voc$ from 920 to \unit[800]{mV}, which is in agreement with previous reports~\cite{Liu2013,Yang2013,Chen2012}. Conversely, the HTL had little effect on the short circuit current and the fill factor; both current-voltage curves exhibit the typical shape of space-charge-limited collection~\cite{Mihailetchi2005}. Averaged photovoltaic characteristics for both types of devices can be found in the Supplemental Material~\cite{SM}.

\begin{figure}
\centering
\includegraphics{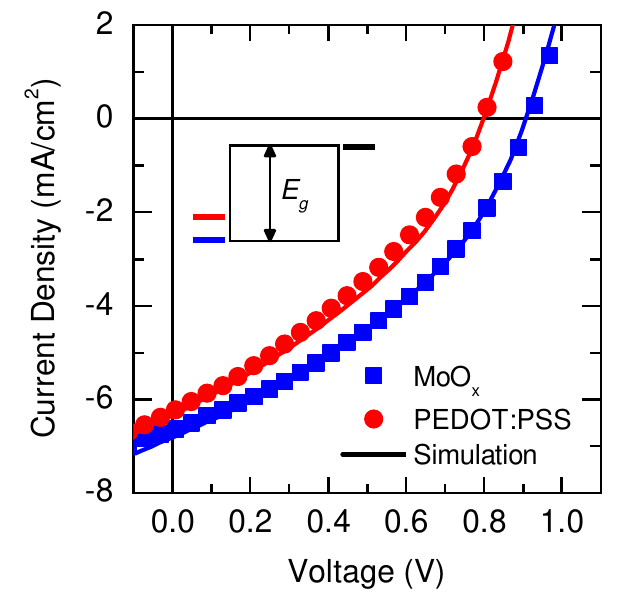}%
\caption{Current-voltage curves for \moox\ (squares) and \pedot\ (circles) devices under simulated AM1.5G~illumination~($\unit[100]{mW/cm^2}$). Solid lines are the result of drift-diffusion simulations using the parameter set in Tab.~\ref{tab:params}. The only parameter varied was the injection barrier $\phian$ at the anode. Inset: Schematic energy level diagram at~$V = \voc$.}
\label{fig:figure01}
\end{figure} 

\subsection{Numerical Model}

We aimed to understand these findings using a numerical drift-diffusion model~\cite{Burgelman2000}. The model treats the bulk heterojunction as an effective semiconductor sandwiched between two electrical contacts. The alignment of the work function of the contacts and the transport levels of the effective semiconductor is given by the injection barriers $\phian$~(anode) and $\phicat$~(cathode). The injection of charge carriers is then assumed to occur via thermionic emission. Surface recombination at the contacts is treated according to Eq.~(\ref{eq:js}).

We then assumed that excess charges are generated by illumination through the transparent anode. To take into account the spatial distribution~$G = G(x)$ of the photogeneration, we coupled the drift-diffusion model with transfer-matrix calculations~\cite{Pettersson1999,Burkhard2010}. The recombination of mobile carriers in the photoactive layer is considered to be solely bimolecular,
\begin{equation}
R = \beta (np - n_i^2),
\label{eq:rec}
\end{equation}
where $n$ and $p$ is the density of electrons and holes, respectively, and $n_i$ the intrinsic carrier density. This is motivated by a recent study~\cite{Scheunemann2017}, where we show that non-geminate recombination SQIB:PCBM blends resembles a second-order process with a prefactor~$\beta$ independent on the carrier density. However, we note that herein, Eq.~(\ref{eq:rec}) is used only as an empirical rate equation, without making any assumptions on the details of the actual recombination mechanism~(e.g., whether it is radiative or non-radiative). All relevant input parameters for the simulation are listed in Tab.~\ref{tab:params}.

\begin{table}[h]
\caption{\label{tab:params}Input parameters used for the numerical drift-diffusion simulations.}
\begin{ruledtabular}
\begin{tabular}{lll}
Parameter & Value & Description\\[.75mm]
\hline
$T$ & \unit[300]{K} & Temperature\\
$\eg$ & \unit[1.36]{eV} & Effective band gap\\
$\varepsilon_r$ & 4 & Dielectric constant\\
$d$ & \unit[100]{nm} & Active-layer thickness\\
$N$ & $\unit[10^{26}]{m^{-3}}$ & Effective density of states\\
$\beta$ & $\unit[10^{-17}]{m^{3}s^{-1}}$ & Recombination coefficent\\
$S$ & $\unit[10^5]{m s^{-1}}$ & Surface recombination velocity\\
$\phicat$ & \unit[0]{eV} & Injection barrier, cathode\\
$\phian$ & varied & Injection barrier, anode\\
$\mun$ & $\unit[2 \times 10^{-8}]{m^2(Vs)^{-1}}$ & Electron mobility\\
$\mup$ & $\unit[2 \times 10^{-10}]{m^2(Vs)^{-1}}$ & Hole mobility\\
\end{tabular}
\end{ruledtabular}
\end{table}

With this model we were able to describe the experimental data only by varying the injection barrier height at the anode, while keeping all other parameters constant~(see solid lines in Fig.~\ref{fig:figure01}). This proves that the variation of the HTL only affects the energy level alignment at the anode, but not the bulk properties of the active layer. Thus, we have at hand a suitable model system to study the effect of imbalanced mobilities on~$\voc$.

\section{Results and Discussion}
\subsection{Impact of an injection barrier}

\begin{figure}
\centering
\includegraphics{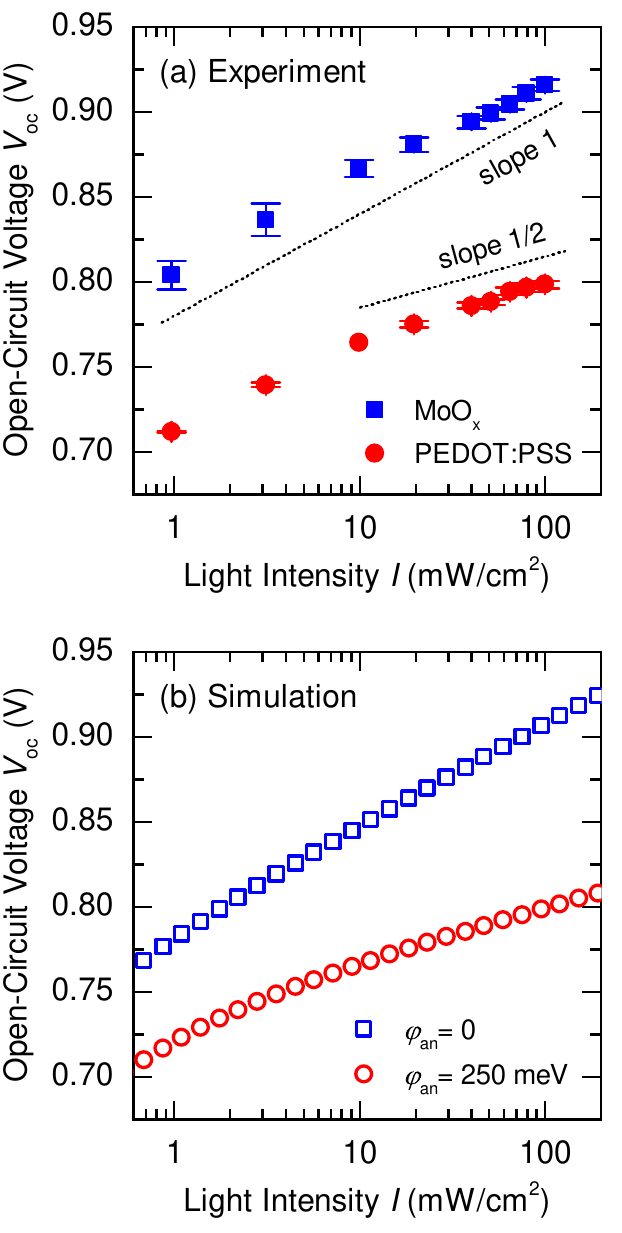}%
\caption{Light-intensity dependence of~$\voc$ under AM1.5G illumination for (a)~experimental and (b)~simulated devices based on \moox~(squares) and PEDOT:PSS~(circles). Error bars in (a) are the standard deviation for 5 individual samples. Dotted lines indicate a scaling of~$kT/q$~(``slope 1") and $kT/2q$~(``slope 1/2").}
\label{fig:figure02}
\end{figure}

Figure~\ref{fig:figure02}(a) shows the experimental light intensity~($I$) dependence of the open circuit voltage. For the \moox\ device, $\voc$ versus $\ln(I)$ has nearly a slope of $kT/q$, as predicted by Eq.~(\ref{eq:koster}) for Ohmic contacts. Hence, we can assume that $\voc$ is limited by bimolecular recombination in the bulk only. In contrast, the \pedot\ device shows a transition towards a lower slope at mid to high light intensity. The reduced slope agrees qualitatively well with Eq.~(\ref{eq:solak}). This kind of behavior, with the slope going from~$\sim kT/q$ at low intensity to~$kT/2q$ at higher intensity~(while $\voc$ remains lower than for the \moox\ device over the entire intensity regime), suggests that surface recombination at one non-Ohmic contact is dominating in the \pedot\ device~\cite{Sandberg2016}.

As can be seen in Fig.~\ref{fig:figure02}(b), both the absolute value and the intensity dependence of $\voc$ are well captured by our numerical model if we only change the magnitude of~$\phian$. It was not possible to reproduce the experimental data by varying the surface recombination velocity at the anode instead~(see Supplemental Material~\cite{SM}). A significant reduction of $S$ would give rise to an extraction barrier, which would then result in S-shaped current-voltage curves~\cite{Sandberg2014,Wilken2014,Wagenpfahl2010,Sundqvist2016}. Because such an S-kink is not present in the data shown in Fig.~\ref{fig:figure01}, we expect the surface recombination current to be mainly determined by the carrier concentrations at the anode and the transport properties of the bulk.

\begin{figure*}
\centering
\includegraphics{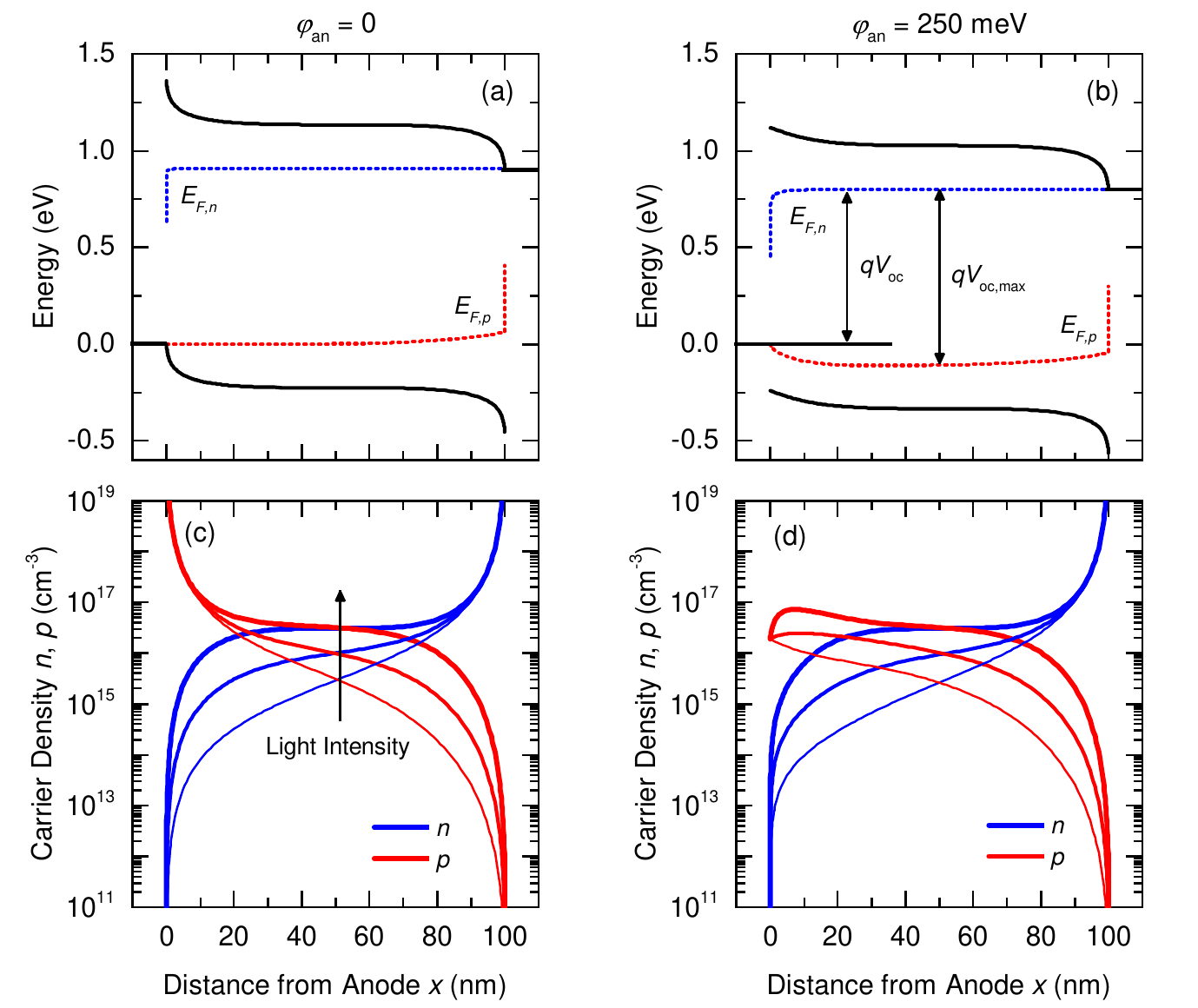}%
\caption{Simulated band diagrams and local carrier concentrations for (a,c)~$\phian = 0$ and (b,d)~$\phian = \unit[250]{meV}$. The anode is positioned at $x = 0$. Solid lines in panels (a) and (b) denote the transport bands and dashed lines the quasi-Fermi levels under AM1.5G illumination~($\unit[100]{mW/cm^2}$). Panels (c) and (d) show the total electron and hole concentration for incident light intensities of 1, 10 and $\unit[100]{mW/cm^2}$. The result of the injection barrier is a gradient of $\efp$ at the anode, which reduces $\voc$ relative to its optimum value $\vocmax$ given by the quasi-Fermi level splitting in the bulk.}
\label{fig:figure03}
\end{figure*}

Figure~\ref{fig:figure03} shows the effect of the anode work function on the energy band diagrams, as well as the electron and hole concentration. The solid lines in panels (a) and (b) denote the transport levels under 1-sun illumination and the dashed lines the quasi-Fermi levels. The \moox\ device shows significant band bending at both electrodes caused by injection of majority carriers into the semiconductor. In case of the anode, there is a high concentration of holes, exceeding the concentration of photogenerated carriers in the bulk by several orders of magnitude. Because of the high hole concentration, electrons are likely to recombine within the bulk, rather than being extracted.  The quasi-Fermi levels at both electrodes are flat, so that the open-circuit voltage represents the splitting of the quasi-Fermi levels in the bulk. The situation remains relatively unchanged with increasing photogeneration. Hence, the light intensity dependence of $\voc$ shows a constant slope over the intensity range studied herein and can be described by bimolecular recombination in the bulk.

For the \pedot\ device with a non-Ohmic contact, the concentration of injected holes is much lower, leading to a reduced band bending at the anode. Consequently, the electron concentration close to the anode is higher than for the case with an Ohmic contact. According to Eq.~(\ref{eq:js}), this non-negligible concentration of minority carriers induces a surface recombination current $\js$. To ensure open-circuit conditions~(no net current), it must be compensated by a hole current
\begin{equation}
j_p = p \mup \frac{d \efp}{d x}.
\end{equation}

Because the magnitude of $p$ close the anode is fixed by the barrier height $\phian$, and $\mup$ is considered constant, an increase of $\js$ due to increasing photogeneration can only be compensated by a gradient of the quasi-Fermi level for holes. At 1-sun illumination, the gradient in $\efp$ is clearly visible. Consequently, the open-circuit voltage is reduced and no longer a measure of the quasi-Fermi level splitting in the bulk.

\begin{figure}
\centering
\includegraphics{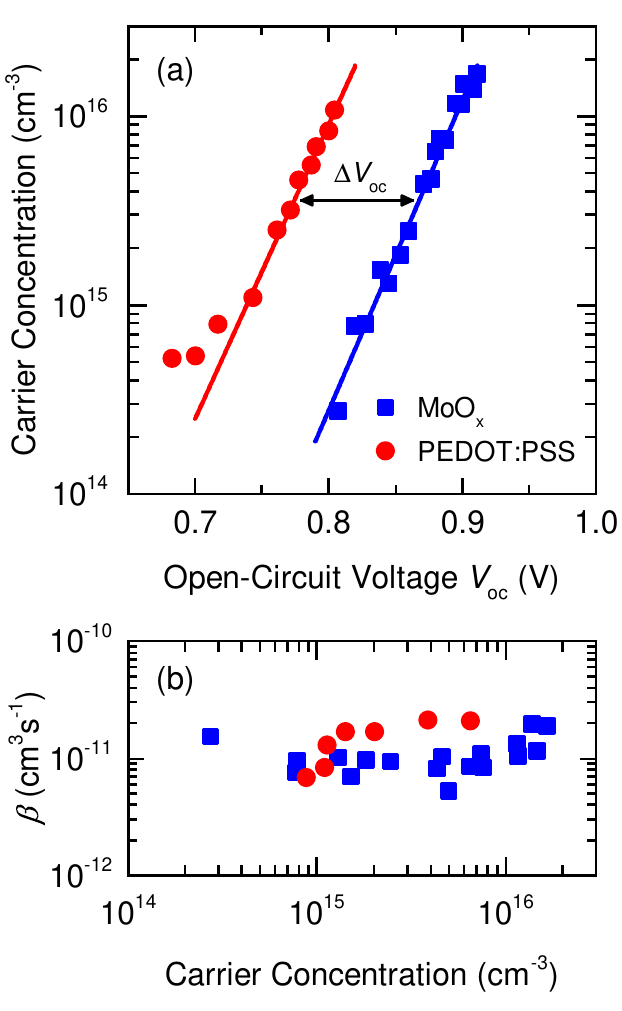}%
\caption{Results of charge extraction and transient photovoltage measurements for \moox~(squares) and \pedot~(circles) devices. (a)~Extracted charge carrier concentration versus open-circuit voltage. (b)~Recombination rate constant~$\beta$ versus carrier concentration.}
\label{fig:figure04}
\end{figure}

Notably, the drift-diffusion model predicts that beyond a thin region of approximately~\unit[15]{nm} close to the anode, both the quasi-Fermi level splitting and the carrier concentrations remain unchanged regardless of the anode work function. To check this prediction, we measured the carrier concentration under open-circuit conditions using bias-assisted charge extraction~\cite{Kniepert2014,Scheunemann2017}. Figure~\ref{fig:figure04}(a) shows indeed only a constant voltage shift~$\Delta\voc$ between the data points for the \moox\ and the \pedot\ sample, while the carrier concentration at a given light intensity remains unchanged.

Furthermore, we performed transient photovoltage experiments to determine the carrier lifetime. By plotting the lifetime versus the carrier concentration (see Supplemental Material~\cite{SM}) we obtained a reaction order close to 2 in both cases. This indicates that the recombination in the bulk is bimolecular, independent of whether an Ohmic or an non-Ohmic hole contact is present. That the nature of the contact does not affect the recombination in the bulk is also evident from Fig.~\ref{fig:figure04}(b), where we plot the recombination rate constant $\beta$ as a function of the carrier concentration. For both devices we find fairly similar values of $\beta \sim \unit[10^{-11}]{cm^3s^{-1}}$.

Hence, we can conclude that the variation in $\voc$ between the \moox\ and \pedot\ devices is solely related to a gradient in the hole quasi-Fermi level at the anode, while the recombination in the bulk is largely unaffected. This is in line with previous studies on material combinations with balanced mobilities~\cite{Wheeler2015, Spies2017}. However, it seems surprising that a relatively low injection barrier of \unit[250]{meV} has such a strong effect on both the magnitude and light intensity dependence of~$\voc$. In the following we show that this is a direct consequence of the highly imbalanced mobilities. 

\subsection{Effect of imbalanced charge transport on the open-circuit voltage}

\begin{figure}
\centering
\includegraphics{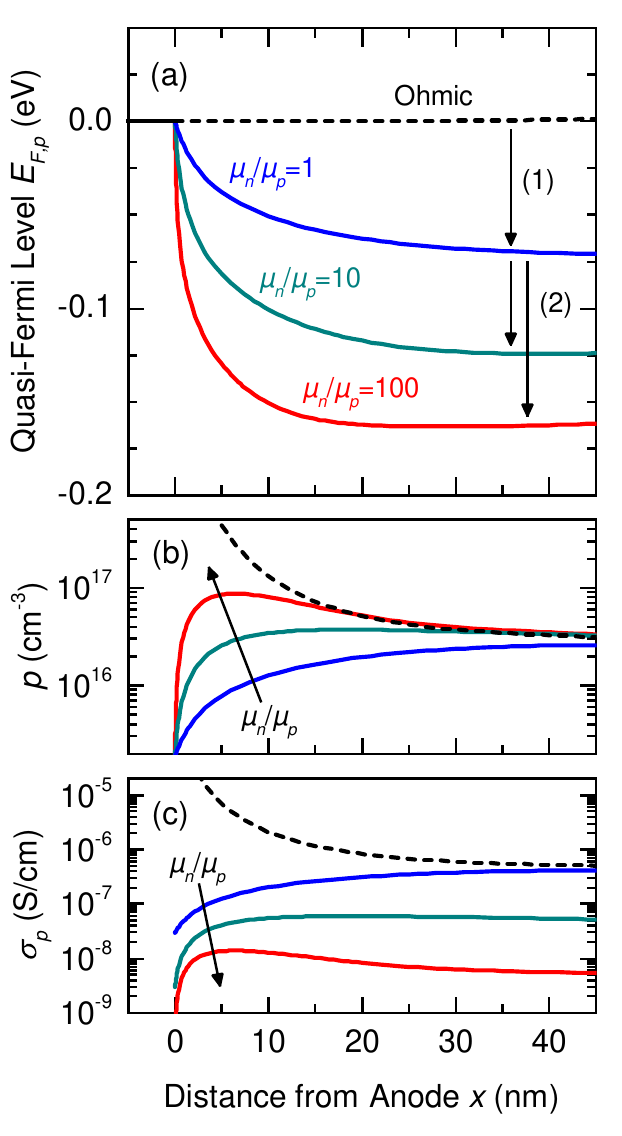}%
\caption{(a) Quasi-Fermi level~$\efp$ close to the anode~(located at $x = 0$) for an ideal device with Ohmic contacts and balanced transport~(dashed line), and for a device with an injection barrier and varied $\mun/\mup$~(solid lines). The arrows indicate the voltage loss due to the barrier height~(1) and due to the imbalanced mobilities~(2), respectively. Panels (b) and (c) show the corresponding  hole concentration~$p$ and hole conductivity~$\sigmap = q \mup p$.}
\label{fig:figure05}
\end{figure}

Having shown that our numerical model describes the experimental data well, we will now use it to discuss the effect of charge transport in more detail. Figure~\ref{fig:figure05}(a) demonstrates how an injection barrier at the anode (similar to the \pedot\ device) affects the hole quasi-Fermi level for different ratios between $\mun$ and $\mup$. For balanced mobilities~($\mun = \mup$), the injection barrier induces a certain gradient $d\efp/dx$, which leads to a voltage loss $\Delta \vocone$ compared to the case with an Ohmic hole contact. If we now lower the hole mobility by one or two orders of magnitude, the gradient of the quasi-Fermi level increases significantly. This can be reflected by introducing a second loss component $\Delta \voctwo$ due to the imbalanced charge transport. Hence, the total loss in $\voc$ can be expressed as
\begin{equation}
\Delta\voc = \Delta\vocone + \Delta\voctwo.
\label{eq:deltavoc}
\end{equation}

Another possible loss mechanism would be a reduction of the quasi-Fermi level splitting (and, thus, the carrier concentration) in the bulk due to very strong surface recombination~\cite{Spies2017}. However, such a reduction is not present here, which is both evident from the bulk recombination measurements~(see Fig.~\ref{fig:figure04}) and the additional band diagrams shown in the Supplemental Material~\cite{SM}. 

Figure~\ref{fig:figure05}(b) and \ref{fig:figure05}(c) illustrate the effect of imbalanced charge transport in more detail. In Fig.~\ref{fig:figure05}(b), an accumulation of holes close to the anode for $\mun/\mup \gg 1$ is clearly seen. However, at the same time, the absolute value of~$\mup$ is decreased. Hence, it is worthwhile to take a look at the conductivity~$\sigmap = q \mup p$. Figure~\ref{fig:figure05}(c) shows that the increase of the hole concentration is not large enough to compensate the decrease of the hole mobility. At the same time, the conductivity~$\sigman$ for electrons is nearly unaffected. Hence, assuming more imbalanced charge transport effectively decreases the difference between $\sigmap$ and $\sigman$ close to the anode. This can be understood in terms of a further loss of selectivity or an virtual increase of the injection barrier height~\cite{Spies2017,Reinhardt2014}.

\begin{figure}
\centering
\includegraphics{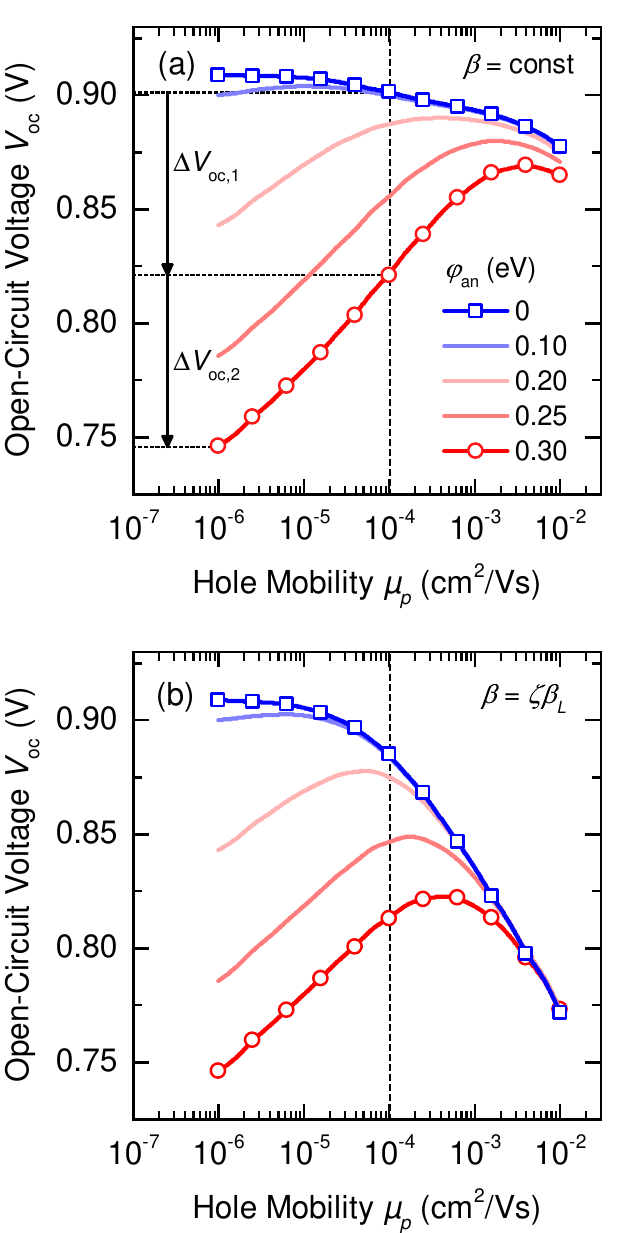}%
\caption{Simulated~$\voc$ for a varied hole mobility but fixed electron mobility of~$\mun = \unit[10^{-4}]{cm^2/Vs}$ and different barrier heights~$\phian$ at the anode. Panel~(a) shows the case of a constant~$\beta$, while for (b) we assumed reduced Langevin recombination with $\beta = \zeta\beta_L$. Vertical dashed lines indicate the case~$\mun = \mup$. The arrows in (a) exemplify the voltage loss~$\Delta \vocone$ for a barrier of~\unit[300]{meV} and $\Delta \voctwo$ for a mobility mismatch of $\mun/\mup = 100$, respectively.}
\label{fig:figure06}
\end{figure} 

Figure~\ref{fig:figure06} shows $\voc$ as a function of $\mup$ for a fixed electron mobility of~$\mun = \unit[10^{-4}]{cm^2/Vs}$ and different barrier heights~$\phian$. First, we discuss the case of a constant~$\beta$, which means the mobility does not affect the bulk recombination~[Fig.~\ref{fig:figure06}(a)]. Only a weak variation of $\voc$ with $\mup$ is then visible for an Ohmic contact~($\phian = 0$). The slight decline of $\voc$ is due to the fact that also the cathode is assumed to be non-selective. When $\mup$ is very large~($\mup \gg \mun$), the device is dominated by hole transport, so that surface recombination at the cathode becomes limiting. If we now introduce a significant barrier at the anode~$(\phian \gg kT$), it is clearly seen that the voltage loss~$\Delta\voc$ becomes determined by the mobility ratio. For~$\mup = \mun$, the reduction of $\voc$ is solely caused by $\Delta \vocone$, which is proportional to~$\phian$. This no longer holds true for imbalanced mobilities. In the case~$\mup \ll \mun$, the voltage loss is further increased by the mobility-dependent~$\Delta \voctwo$, and a logarithmic dependence of $\voc$ on $\mup$ can be seen. In contrast, for $\mup \gg \mun$, surface recombination is partly compensated, as charges accumulate now at the (Ohmic) cathode. As a result, the total voltage loss is effectively reduced~($\Delta \voctwo < 0$). We also did simulations for $d = 70$ and \unit[140]{nm}~(see Supplemental Material~\cite{SM}). Previously, we have shown that this thickness range produces clear differences in the competition between charge extraction and bimolecular recombination~\cite{Scheunemann2017}. However, we find here that the mobility dependence of~$\voc$ is fairly unaffected by the active-layer thickness. This shows that our results are independent on the collection of majority carriers.

Next, we consider the case that also the bulk recombination is limited by diffusion~[Fig.~\ref{fig:figure06}(b)]. Such a process is commonly described by the Langevin model, predicting a mobility-dependent recombination coefficient
\begin{equation}
\beta_L = \frac{q}{\varepsilon\varepsilon_0}(\mun + \mup).
\label{eq:langevin}
\end{equation}
However, it is known that the recombination in phase-separated organic blends is reduced compared to the Langevin model, $\beta = \zeta \beta_L$, where $\zeta$ is a reduction factor~\cite{Pivrikas2005,Murthy2013,Gohler2018}. Here, we chose~$\zeta = 0.1$, so that with the given mobilities of the SQIB:PCBM system, the experimental value of $\beta$ is reproduced. For $\mup \ll \mun$, no difference in the mobility dependence of $\voc$ can be seen between Fig.~\ref{fig:figure06}(a) and Fig.~\ref{fig:figure06}(b), as $\beta$ is largely determined by the fixed~$\mun$. If the mobilities are similar, there is a significant contribution of $\mup$ to the magnitude of $\beta$. In the case $\mup \gg \mun$, the coefficient~$\beta$ becomes so large that the device is entirely dominated by bulk recombination, and the barrier at the anode is no longer relevant. Altogether, this results in a maximum of $\voc(\mup)$, which shifts towards larger values of~$\mup$ with increasing~$\phian$. Such an optimum value of $\voc$ is unique for imbalanced charge transport and was not observed in previous studies, where $\mun$ and $\mup$ were varied simultaneously~\cite{Tress2012,Kirchartz2009,Wagenpfahl2010b}. 

Our numerical results show that independent of the bulk recombination mechanism, surface recombination at the anode critically determines $\voc$ for imbalanced mobilities with $\mup \ll \mun$. In contrast, for $\mup \gg \mun$, the quality of the anode is less important. We note that our conclusions are directly transferable to the case of a non-Ohmic cathode. A significant barrier~$\phicat$ would severely limit $\voc$ for $\mun \ll \mup$, which is a relevant scenario for non-fullerene solar cells~\cite{Holliday2016,Yan2018,Hou2018}. 

\subsection{Analytical expression for $\voc$ in the case of imbalanced mobilities}
Recently, Sandberg~et~al.~\cite{Sandberg2016} provided analytical means to describe~$\voc$ for different cases related to surface recombination. If one contact is non-Ohmic~(here the anode) and surface recombination is limited by the diffusion of minority carriers~(here electrons; effective velocity~$v_{d,n}$) rather than the interface kinetics~($S \gg v_{d,n}$), the authors derived the expression
\begin{equation}
q\voc = \eg - \phian - kT \ln\left(\frac{v_{d,n} N}{Gd}\right)
\label{eq:sandberg1}
\end{equation}
for low light intensities, where bulk recombination is negligible. When~$\voc$ is far from flat-band conditions, we have~$v_{d,n} \approx \mun |F(0)|$, where $F(0)$ is the electric field close to the anode. Equation~(\ref{eq:sandberg1}) is valid as long as $L_p^\ast \gg d$, with the effective diffusion length 
\begin{equation}
L_p^\ast \approx \left(\frac{\mueff k T}{q \sqrt{\beta G}}\right)^{1/2},
\label{eq:ldiff}
\end{equation}
where~$\mueff = 2\sqrt{\mun\mup}$ is an effective mobility. At high enough light intensities, so that $L_p^\ast \ll d$ and flat-band conditions prevail on the anode side of the active layer, the surface recombination is restricted to a region given by the effective diffusion length. Under these conditions, $\voc$ can be approximated by~\cite{Sandberg2014,Sandberg2016}
\begin{equation}
q\voc \approx \eg - \phian - \frac{kT}{2} \ln\left(\frac{\mun \beta N^2}{\mup G}\right),
\label{eq:sandberg2}
\end{equation}
which in the limit~$\mun \rightarrow \mup$ is equivalent to the result of Solak~et~al.~\cite{Solak2016}. We note that Eq.~(\ref{eq:sandberg2}) already provides a framework to predict~$\voc$ for moderate mobility contrasts; however, its derivation assumes that flat-band conditions prevail close to the surface recombination dominated region near the anode. This is no longer valid for for highly imbalanced mobilities.

\begin{figure}
\centering
\includegraphics{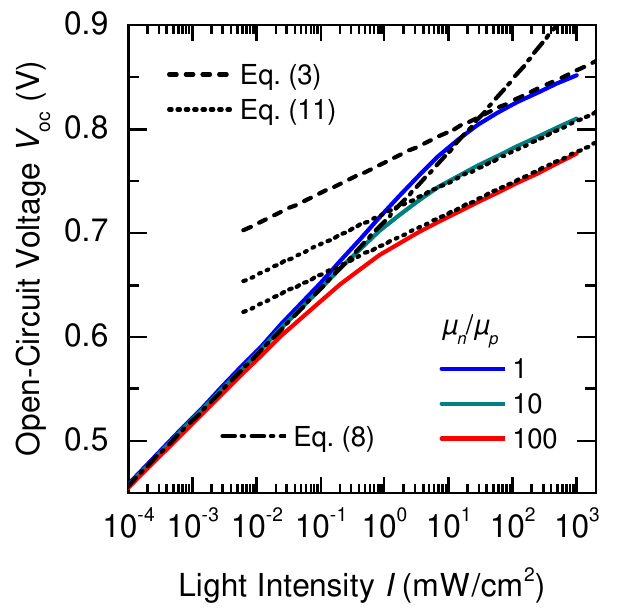}%
\caption{Simulated~$\voc$~(solid lines) versus light intensity for a device having a non-Ohmic anode~($\phian=\unit[0.3]{eV}$) at different ratios between~$\mun$ and~$\mup$. Equations~(\ref{eq:solak}) and (\ref{eq:sandberg3}) are indicated by dashed and  dotted lines, respectively. Our extended analytical expression Eq.~(\ref{eq:sandberg3}) shows striking agreement with the simulations for highly imbalanced mobilities. The dash-dotted line indicates Eq.~(\ref{eq:sandberg1}), valid in the low-intensity regime, with~$v_{d,n}$ approximated according to Ref.~\citenum{Sandberg2016}.}
\label{fig:figure07}
\end{figure}

For~$\mup\ll\mun$, a considerable pile-up of holes is taking place close to the anode at high light intensities, as is clearly visible in Fig.~\ref{fig:figure03}(d). As evident from Fig.~\ref{fig:figure03}(b), the resulting space charge region of holes gives rise to a photo-induced upward energy-level bending near the anode. In the hole-dominated space charge region, the bulk recombination of holes is negligibly small. Instead, at open-circuit conditions, the hole current within this region is balanced by an equal, but opposite surface recombination current of electrons, diffusing against the band bending at the anode. After accounting for the hole-induced energy-level bending and the associated electron diffusion, Eq.~(\ref{eq:sandberg2}) modifies to
\begin{equation}
q\voc = \eg - \phian - \frac{kT}{2} \ln\left(\frac{q \mun^2 N^2}{\varepsilon\varepsilon_0\mup G}\right),
\label{eq:sandberg3}
\end{equation}
as shown in Supplemental Material~\cite{SM}. The main feature of Eq.~(\ref{eq:sandberg3}) is that it no longer depends on the bimolecular recombination strength~$\beta$. Hence, in the limit of highly imbalanced mobilities and at high enough light intensities, $\voc$ becomes independent of bulk recombination and is solely given by the contacts and the charge transport in the active layer. Equation~(\ref{eq:sandberg3}) also explains the logarithmic $\mup$ dependence at high mobility contrasts~$\mun/\mup \gg 1$ seen in Fig.~\ref{fig:figure06}. The voltage loss due to the imbalanced mobilities is then given by
\begin{equation}
\Delta \voctwo = \frac{kT}{2} \ln\left(\frac{q \mun^2}{\varepsilon\varepsilon_0\mup \beta}\right).
\label{eq:sandberg4}
\end{equation}

As shown in Fig.~\ref{fig:figure07} for mobility contrasts of one and two orders of magnitude, the modified analytical expression in Eq.~(\ref{eq:sandberg3}) describes the numerical data well in the high-intensity regime. We note that as long as the anode can be considered as non-Ohmic~($\phian \gg kT$), this holds true also for other injection barrier heights~(see Supplemental Material~\cite{SM}). In the low-intensity regime, $\voc$ becomes independent of the mobility contrast and can be described by Eq.~(\ref{eq:sandberg1}) instead. Our analysis demonstrates that the transition is shifted towards lower photogeneration for increasing~$\mun/\mup$.

Furthermore, we checked the validity of our analytical framework, as detailed in the Supplemental Material~\cite{SM}. We find that Eq.~(\ref{eq:sandberg3}) predicts the open-circuit voltage for mobility contrasts~$\mun/\mup \geq 5$ with an relative error below~1\%~(for the parameters in Tab.~\ref{tab:params}). It is worth noting that Eq.~(\ref{eq:sandberg2}), which neglects band bending in the hole-induced space charge region, reaches a similar accuracy only for a mobility imbalance of less than a factor of~2. Finally, we point out that Eq.~(\ref{eq:koster}) is in none of the cases presented in Fig.~\ref{fig:figure07} suitable to describe the data, even in the low-intensity regime and even though only pure bimolecular recombination was assumed in the simulation. Hence, special care has to be taken when trying to assess information about the bulk recombination from the slope of~$\voc$ versus $\ln(I)$, also called the light ideality factor~\cite{Kirchartz2013}. If the contacts are not sufficiently selective, the ideality factor will always be affected by surface recombination.

\section{Conclusion}
In summary, we have studied how imbalanced charge transport affects the interplay of bulk and surface recombination in organic solar cells. Combining experiments and simulations for a blend system with a strong mobility mismatch, we have identified two cases with respect to the energy level alignment at the electrodes: For Ohmic contacts, the open-circuit voltage $\voc$ still is representative of the quasi-Fermi level splitting in the bulk, even though the mobilities $\mun$ and $\mup$ are highly imbalanced. However, if one contact is non-Ohmic, $\voc$ becomes critically determined by the mobility ratio. For the devices studied herein~($\mun/\mup \gg 1$), we find that surface recombination at the anode reduces~$\voc$ more strongly than it would be the case with balanced mobilities. The reason is that with decreasing $\mup$, a larger gradient of the quasi-Fermi level is required to cancel out the surface recombination current of electrons. An analogous situation occurs for a device dominated by hole transport~($\mun/\mup \ll 1$) at a the cathode. Hence, it is properties of the photoactive blend that decide whether an electrode can be considered appropriate.

We have also derived analytical equations for~$\voc$ that take into account the pile-up of space charge due to highly imbalanced mobilities. In particular, Eq.~(\ref{eq:sandberg3}) shows excellent agreement with the data from our experimentally validated numerical device model. With this we hope to provide a framework that helps researchers in designing efficient organic photovoltaics from materials with imbalanced charge transport. 

\begin{acknowledgements}
The authors thank Matthias Schulz and Arne L\"{u}tzen~(University of Bonn, Germany) for providing the squaraine dye, as well as Mathias Nyman and Staffan Dahlstr\"{o}m~(\AA{}bo Akademi University) for fruitful discussions. D.\,S., S.\,W. and M.\,S. thank J\"{u}rgen Parisi for constant support of their research in Oldenburg. Financial support by the Magnus Ehrnrooth Foundation and the Research Mobility Programme of \AA{}bo Akademi University is gratefully acknowledged. 
\end{acknowledgements}

\bibliography{references}

\end{document}

\title{Supplemental Material for ``Effect of Imbalanced Charge Transport on the Interplay of Surface and Bulk Recombination in Organic Solar Cells"}

\author{Dorothea Scheunemann}
\email[]{dorothea.scheunemann@uol.de}
\author{Sebastian Wilken}
\affiliation{Physics, Faculty of Science and Engineering and Center for Functional Materials, \AA{}bo Akademi University, Porthansgatan 3, 20500 Turku, Finland}
\affiliation{Institute of Physics, Energy and Semiconductor Research Laboratory, Carl von Ossietzky University of Oldenburg, 26111 Oldenburg, Germany}

\author{Oskar J.~Sandberg}
\affiliation{Department of Physics, Swansea University, Singleton Park, Swansea, SA2 8PP, Wales, United Kingdom}

\author{Ronald \"{O}sterbacka}
\affiliation{Physics, Faculty of Science and Engineering and Center for Functional Materials, \AA{}bo Akademi University, Porthansgatan 3, 20500 Turku, Finland}

\author{Manuela Schiek}
\affiliation{Institute of Physics, Energy and Semiconductor Research Laboratory, Carl von Ossietzky University of Oldenburg, 26111 Oldenburg, Germany}

\maketitle
\tableofcontents

\section*{Determination of the Build-In Voltage}
\begin{figure}[h!]
\centering
\includegraphics{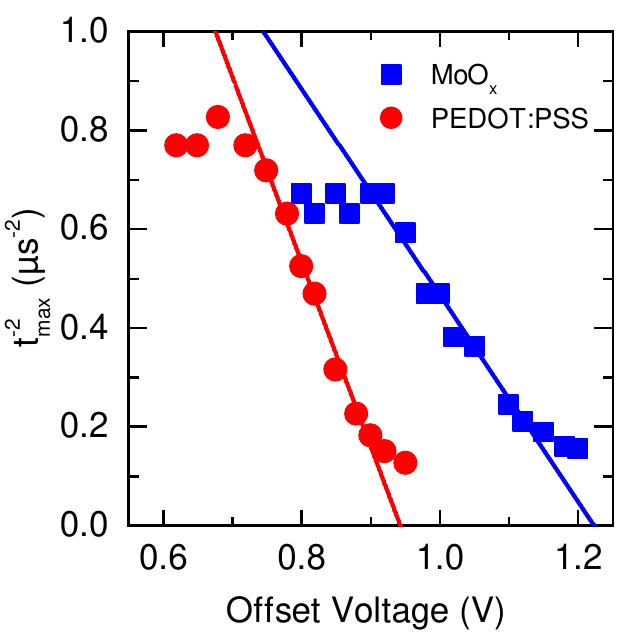}
\caption{Determination of~$\vbi$ for devices with a hole transport layer of \moox~(squares) and \pedot~(circles) from the results of dark CELIV measurements.}
\label{fig:S1}
\end{figure}

To estimate the built-in voltage~$\vbi$, we followed a method that has recently been developed by Dahlstr\"om~et~al.~\cite{Dahlstrom2018}. Therefore, we determined the maximum extraction current time~$\tmax$ from charge carrier extraction by linearly increasing voltage~(CELIV) measurements in the dark and at different offset voltages~$\voff$. As shown in Ref.~\citenum{Dahlstrom2018}, $\tmax$ depends on the voltage according to
\begin{equation}
\tmax^{-2} = \frac{qA\mu}{kTd^2}(\vbi - \voff),
\end{equation}
where $A$ is the slope of the linearly increasing voltage pulse, $\mu$ the charge carrier mobility, $k$ the Boltzmann constant, $T$ the temperature, $d$ the active layer thickness and $q$ the elementary charge. Hence, $\vbi$ can be estimated from a linear fit of~$\tmax^{-2}$ versus $\voff$, see Fig.~\ref{fig:S1}. Prior to the analysis, we corrected the data for $RC$ time effects by replacing $\tmax$ with $\tmax - 3R_\text{s}C_\text{geo}$, where $R_\text{s}$ is the sum of series and load resistance and $C_\text{geo}$ the geometrical capacitance of the device. The intersection of the linear fits with the voltage axis yields  $V_\text{bi} \approx \unit[0.94]{V}$ for the \pedot\ device and $\vbi \approx \unit[1.22]{V}$ for the \moox\ device.

\clearpage
\section*{Photovoltaic Characteristics}
Current--voltage characteristics under AM1.5G illumination were recorded with a semiconductor characterization system (Keithley 4200) and a solar simulator (Photo Emission Tech., class AAA) at ambient conditions. The light intensity was adjusted to $\unit[100]{mW/cm^2}$ using a calibrated silicon solar cell. Spectral mismatch was not taken into account.

\begin{table}[h!]
  \caption{\label{tab:IVParams}Average photovoltaic characteristics of the SQIB:PCBM solar cells with varied HTL under simulated AM1.5G illumination.}
  \begin{ruledtabular}
  \begin{tabular}{ccccc}
    HTL (nm) & $J_{\text{sc}}$ (mA/cm$^2$) & $V_{\text{oc}}$ (V) & FF (\%) & PCE (\%)  \\
    \hline
   MoO$_\text{x}$ & 6.7 $\pm$ 0.1 & 0.92 $\pm$ 0.01 & 38 $\pm$ 1 & 2.30 $\pm$ 0.03\\
   PEDOT:PSS & 6.1 $\pm$ 0.2 & 0.80 $\pm$ 0.01 & 34 $\pm$ 1 & 1.67 $\pm$ 0.04\\
   \end{tabular}
\end{ruledtabular}
\end{table}

\clearpage
\section*{Surface Recombination Velocity}
\begin{figure}[h!]
\centering
\includegraphics{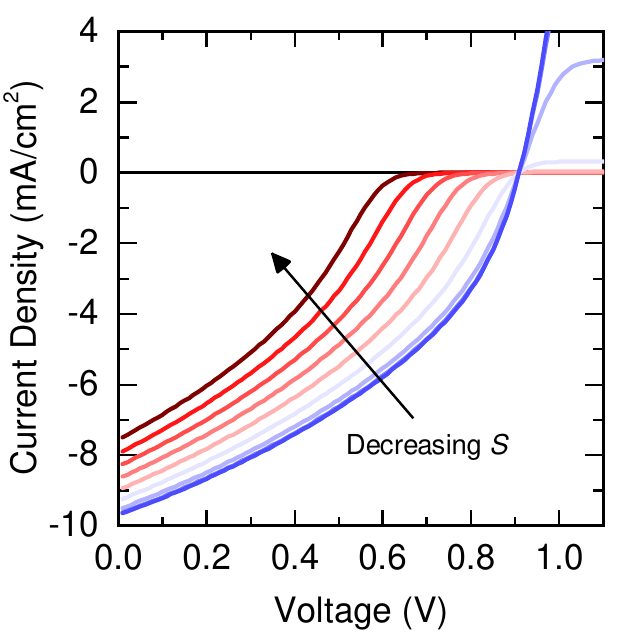}
\caption{Simulated current-voltage curves for a SQIB:PCBM device with varying surface recombination velocity at the anode, but otherwise ideal contact properties ($\phian = \phicat = 0$).}
\label{fig:S2}
\end{figure}

We also attempted to model the differences between the \moox\ and the \pedot\ devices by varying the surface recombination velocity~$S$ at the anode. As can be seen in Fig.~\ref{fig:S2}, a reduction of~$S$ predominantly results in a deformation of the current-voltage curves around~$V = \voc$~(``S-shape"), in accordance with literature~\cite{Wilken2014,Sundqvist2016}. However, such an S-shape behavior was not present in our experimental data.

\clearpage
\section*{Determination of the Reaction Order}
\begin{figure}[h!]
\centering
\includegraphics{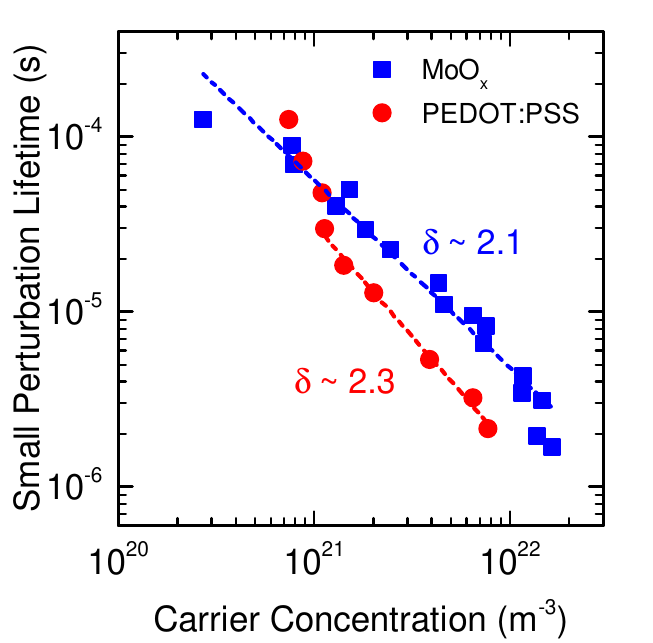}
\caption{Small-perturbation lifetime~$\taudn$ versus carrier concentration~$n$ as obtained from combined TPV and BACE measurements. Dashed lines are fits according to~$\taudn \propto n^{1-\delta}$.}
\label{fig:S3}
\end{figure}

To estimate the reaction order of the bulk recombination, we plotted the small-perturbation lifetime~$\taudn$ from transient photovoltage~(TPV) measurement as a function of the carrier concentration~$n$ from bias-extracted charge extraction~(BACE) measurements under the same experimental conditions~(Figure~\ref{fig:S3}). According to previous studies~\cite{Foertig2014,Wheeler2015}, we applied a power-law fit according to $\taudn \propto n^{1-\delta}$ to these data, which yields an empirical reaction order of~$\delta = 2.1$~(\moox) and $\delta = 2.3$~(\pedot), respectively.

\clearpage
\section*{Additional Band Diagrams}
\begin{figure}[h!]
\centering
\includegraphics{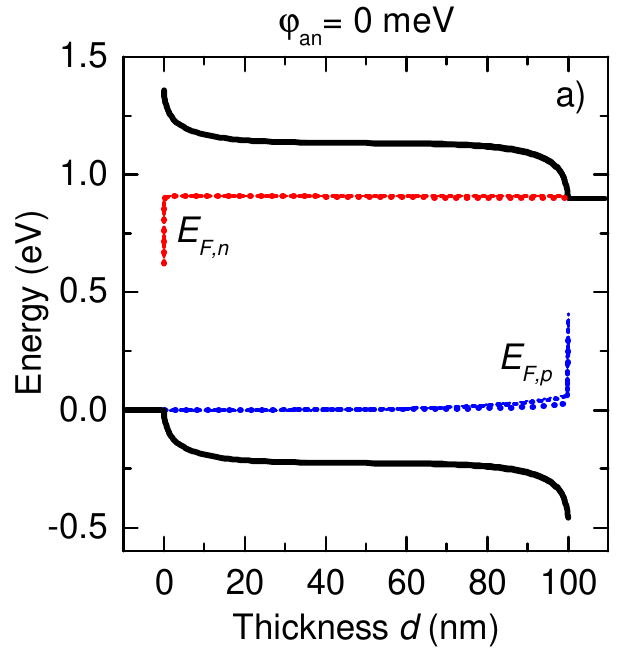}
\hspace{1cm}
\includegraphics{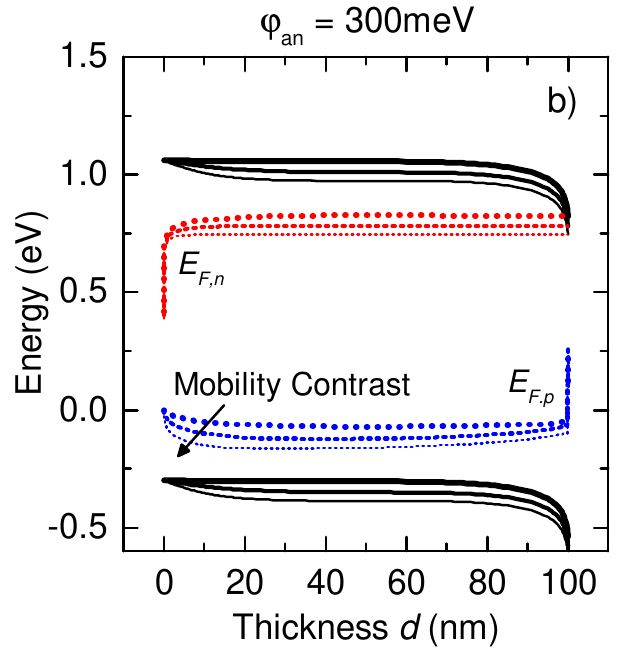}
\caption{Simulated band diagrams at open circuit and 1-sun illumination for an injection barrier height of (a)~$\phian = \unit[0]{mV}$ and (b)~$\varphi_\text{an} = \unit[300]{mV}$ and for mobility contrasts of $\mun/\mup=100$ (thin lines), $\mun/\mup=1$ (medium lines) and $\mun/\mup=0.01$ (thick lines), respectively. The electron mobility was fixed to~$\mun = \unit[10^{-4}]{cm^2/Vs}$.}
\label{fig:S4}
\end{figure}

Figure~\ref{fig:S4} shows band diagrams for different ratios between the electron and hole mobility. It is clearly seen that, independent of the barrier height $\phian$, the mobility contrast has virtually no effect on the quasi-Fermi level splitting \textit{within} the photoactive layer.

\clearpage
\section*{Variation of the Active-Layer Thickness}
\begin{figure}[h!]
\centering
\includegraphics{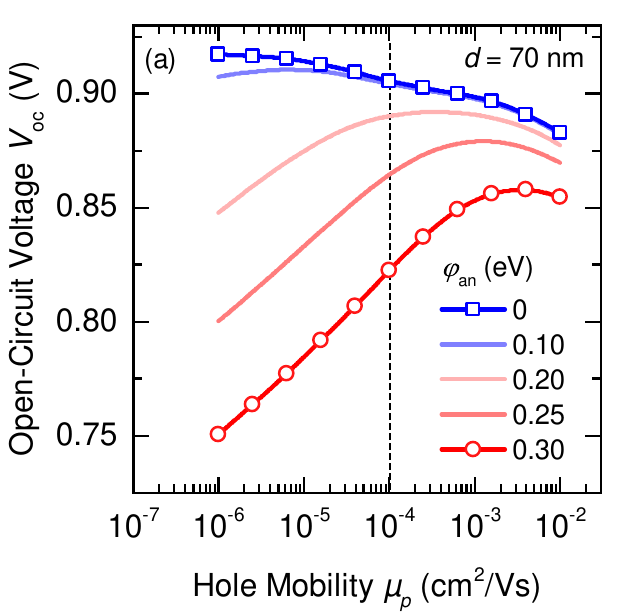}
\hspace{.5cm}
\includegraphics{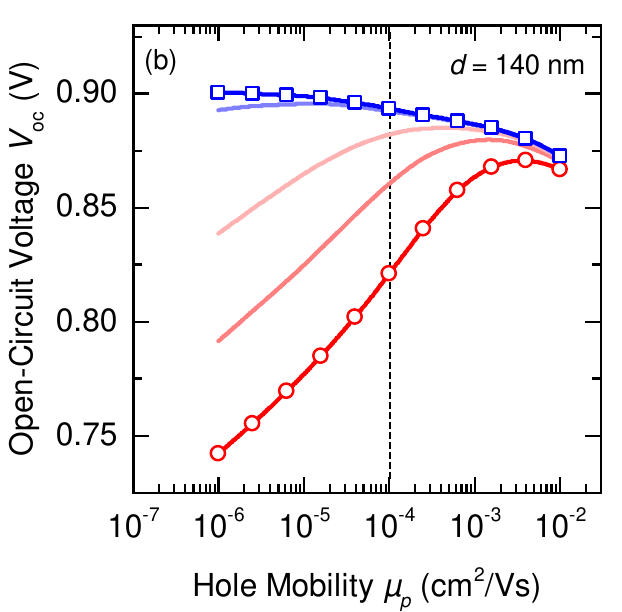}
\caption{Simulated~$\voc$ for a varied hole mobility but fixed electron mobility of~$\mun = \unit[10^{-4}]{cm^2/Vs}$ and different barrier heights~$\phian$ at the anode. Panel~(a) shows the result for an active-layer thickness of~$d = \unit[70]{nm}$, and panel~(b) for an active-layer thickness of~$d = \unit[140]{nm}$.}
\label{fig:SX}
\end{figure}

\clearpage
\section*{Analytical Model}
We assume the electron and hole mobilities to be highly imbalanced, with holes being the slower charge carriers~($\mun/\mup\gg 1$). Under these conditions, provided that the light intensity is high enough, holes will pile up within the active layer; as a result, a hole-dominated mobility-induced space charge region~($0 < x < w_s$) is formed adjacent to the anode, where the thickness of the space charge region is given by~$w_s < d$. The space charge will screen the electric field in the region~($x > w_s$) immediately outside the space charge region. In the region~$x > w_s$~(not too close to the cathode), the electric field is close to zero and the generation rate is closely balanced by recombination. 

Since the space charge region is dominated by holes~($p \gg n$), the bulk recombination of holes is negligible within this region~($0 < x < w_s$). Assuming the hole current to be drift-dominated~(i.e., the electric field is strong enough), the continuity equation, the hole current equation and the Poisson equation, respectively, in the region $0 < x < w_s$ read:
\begin{equation}
\frac{1}{q} \frac{\mathrm{d}j_p}{\mathrm{d}x} = G,
\label{eq:ana:set1}
\end{equation}
\begin{equation}
j_p(x) = q \mup p(x) F(x),
\label{eq:ana:set2}
\end{equation}
\begin{equation}
\frac{\mathrm{d}F}{\mathrm{d}x} = \frac{q}{\varepsilon\varepsilon_0} p(x).
\label{eq:ana:set3}
\end{equation}
Then, assuming $j_p(w_s) = F(w_s) = 0$, and taking the photogeneration rate~$G$ to be relatively homogenous within the space charge region, Eqs.~(\ref{eq:ana:set1}) to~(\ref{eq:ana:set3}) can be combined as 
\begin{equation}
j_p(x) = -q G (w_s - x) = \mup \varepsilon\varepsilon_0 F(x) \frac{\mathrm{d}F}{\mathrm{d}x}
\end{equation}
for $x < w_s$ and solved for the electric field,
\begin{equation}
F(x) = \sqrt{\frac{qG}{\mup \varepsilon\varepsilon_0}} (x - w_s).
\label{eq:ana:efield}
\end{equation}
We note that this analysis yields the correct Goodman and Rose~\cite{Goodman1971} type voltage and generation dependence expected for space-charge-limited photocurrents,
\begin{equation}
\jph \approx qGw_s \propto G^{3/4} \sqrt{V_0-V},
\end{equation}
as discussed in Ref.~\citenum{Sandberg2018}.

As the bulk recombination is assumed negligibly small within the space charge region, at open-circuit conditions the hole current is exactly balanced by an equal, but opposite surface recombination current of electrons, so that $j = j_p(x) + j_n(x) = 0$. Subsequently, $j_p(0) = -j_n(0)$ must hold true at~$\voc$. A general expression for the surface recombination current $j_n(0)$ of electrons at the anode was derived by Sandberg~et~al.~\cite{Sandberg2016} from the drift-diffusion equations. Equation~(A4) from Ref.~\citenum{Sandberg2016} can be readily rewritten as
\begin{equation}
\int_0^d j_n(x) \exp \left[\frac{E_c(x)-E_c(0)}{kT}\right]\mathrm{d}x = \mun kT \nan\left[\exp\left(\frac{qV}{kT}\right)-1\right],
\label{eq:ana:intergal}
\end{equation}
whereby it is assumed that the electron transport is limited by diffusion rather than by the interface kinetics at the anode. Here, $\nan = N \exp\left[-(\eg-\phian)/kT\right]$ is the equilibrium electron density at the anode. Provided that the bulk recombination within the space charge region remains negligible, we may write the electron current as
\begin{equation}
-\frac{1}{q} \frac{\mathrm{d}j_n}{\mathrm{d}x} = G
\end{equation}
for $x < w_s$, which then yields
\begin{equation}
j_n(x) = j_n(0) - qGx.
\end{equation}
Since the main contribution to the integral in Eq.~(\ref{eq:ana:intergal}) is from the region close to $x = 0$, we can approximate $E_c(x) \approx E_c(0) + qF(0)x$ and rewrite Eq.~(\ref{eq:ana:intergal}) as
\begin{equation}
j_n(0) \approx -q \mun F(0) \nan \left[\exp\left(\frac{qV}{kT}\right)-1\right]
\label{eq:ana:elcurrent}
\end{equation}
under the assumption $q |F(0)| w_s\gg kT$. At open circuit~($V = \voc$), where $j_p(0) = -j_n(0) = - q G w_s$, it then follows directly form Eq.~(\ref{eq:ana:efield}) and Eq.~(\ref{eq:ana:elcurrent}) that
\begin{equation}
\exp\left(\frac{q\voc}{kT}\right) \approx - \frac{Gw_s}{\mun F(0)n_{an}} = \frac{1}{\nan}\sqrt{\frac{\varepsilon\varepsilon_0\mup G}{q\mun^2}}
\end{equation}
assuming $q\voc\gg kT$. Using the definition of $\nan$, we finally arrive at
\begin{equation}
q\voc = \eg - \phian - \frac{kT}{2}\ln\left(\frac{q\mun^2N^2}{\varepsilon\varepsilon_0\mup G}\right).
\end{equation}
Comparing this expression with Eq.~(3) in the main text, the enhanced surface recombination loss in the open-circuit voltage due to the mobility imbalance is given by 
\begin{equation}
\Delta \voctwo = \frac{kT}{2} \ln\left(\frac{q \mun^2}{\varepsilon\varepsilon_0\mup \beta}\right).
\end{equation}

\clearpage
\section*{Validity of Analytical Model for Different Injection Barrier Heights}
\begin{figure}[h!]
\centering
\includegraphics{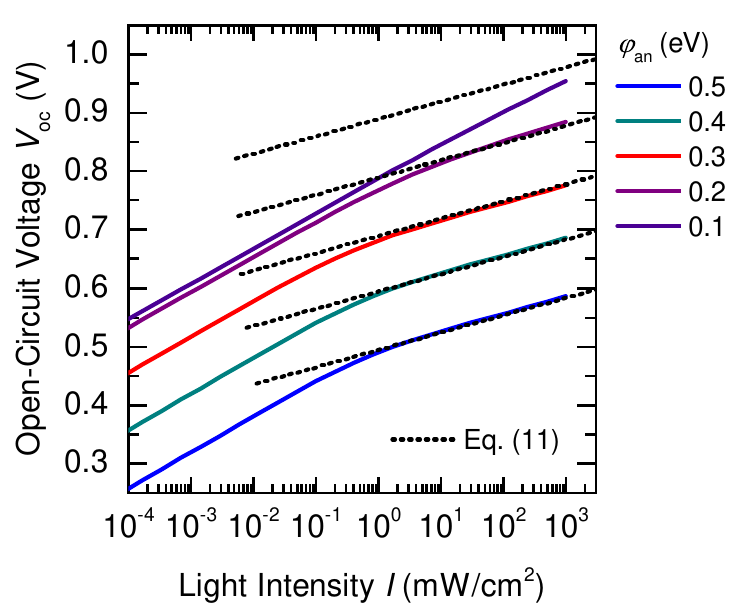}
\caption{Simulated open-circuit voltage~(solid lines) versus light intensity for a device with a mobility contrast of~$\mun/\mup=100$ and different injection barrier heights~$\phian$ at the anode. Dotted lines are the prediction of Eq.~(11) in the main text.}
\end{figure}

\clearpage
\section*{Validity and Limits of Analytical Equations}
\begin{figure}[h!]
\centering
\includegraphics{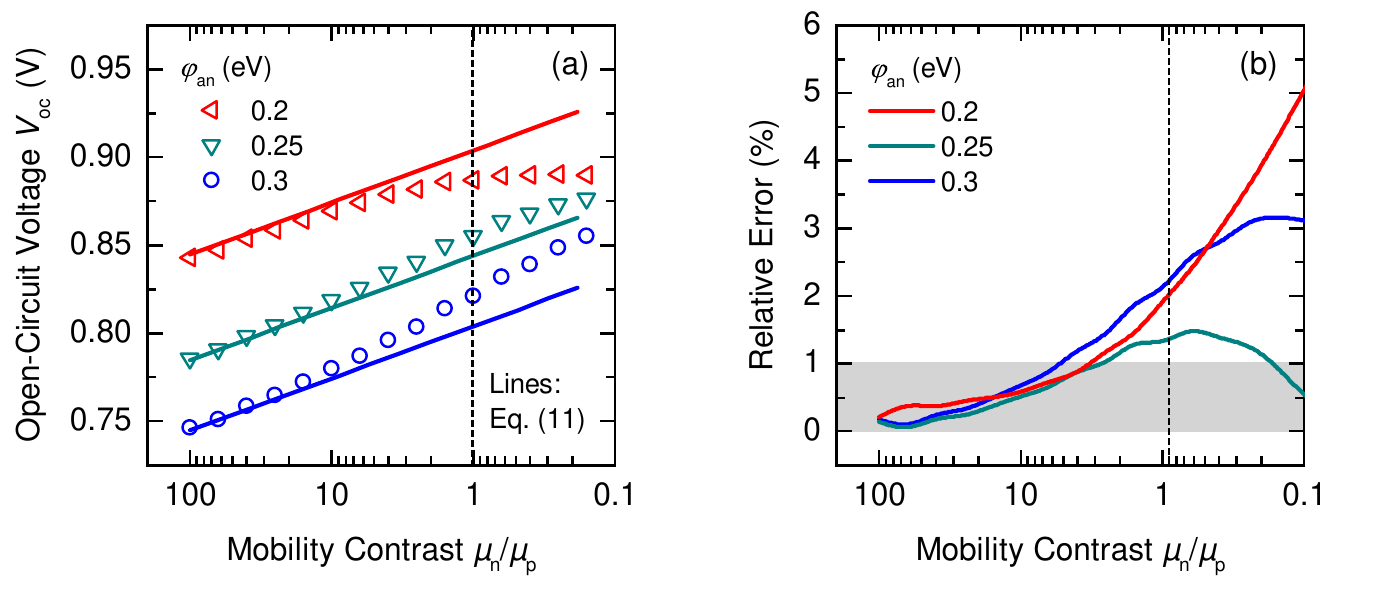}
\includegraphics{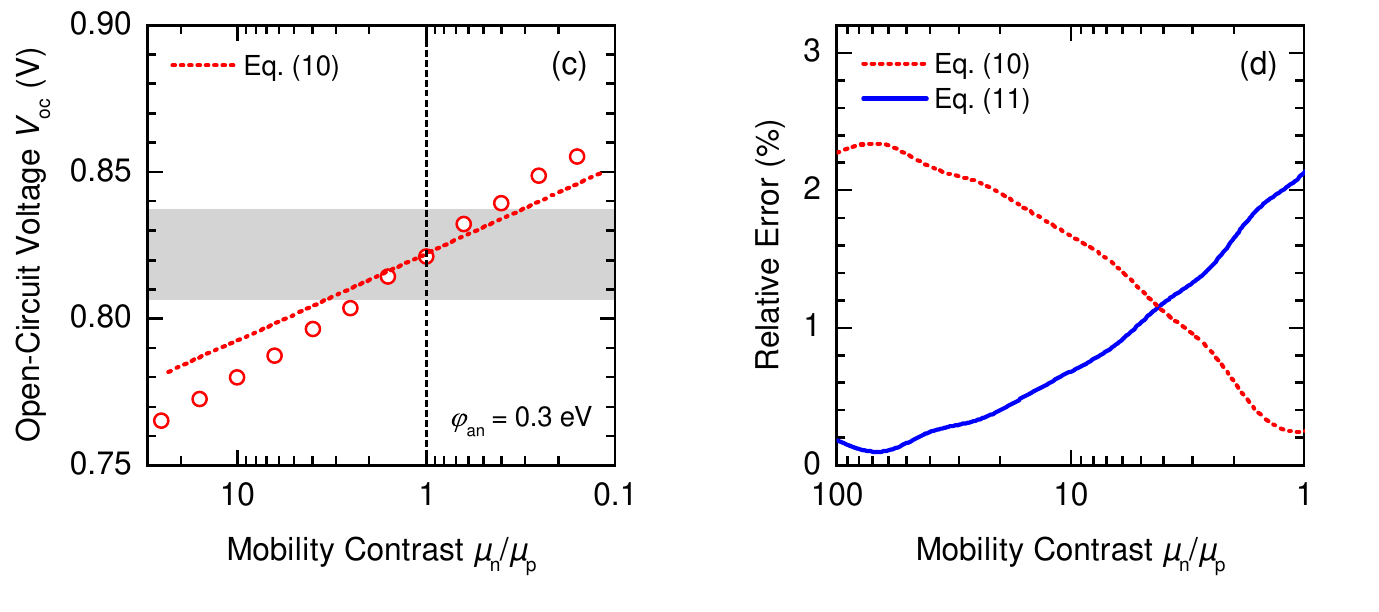}
\caption{(a) Simulated dependence of $\voc$ on the mobility contrast $\mun/\mup$ for different injection barrier heights~$\phian$ at the anode. Solid lines are the prediction of Eq.~(11) in the main text. (b)~Relative error between numerical and analytical data. Shadowed area corresponds the the region with $< 1\%$ error. (c)~Numerical data for an injection barrier of~$\phian = \unit[0.3]{eV}$ together with the prediction of Eq.~(10) in the main text. Shadowed area indicates the region with $< 1\%$ error between simulation and analytical model. (d)~Comparison of the relative error produced by Eq.~(10) and Eq.~(11), respectively, for a device with $\phian = \unit[0.3]{eV}$.}
\end{figure}

\bibliography{references_SM}